\documentclass{Interspeech2024}

\interspeechcameraready 

\usepackage[utf8]{inputenc} 
\usepackage[T1]{fontenc}    
\usepackage{hyperref}       
\usepackage{url}            
\usepackage{booktabs}       
\usepackage{amsfonts}       
\usepackage{nicefrac}       
\usepackage{microtype}      
\usepackage{xcolor}         
\usepackage{cite} 
\hypersetup{colorlinks=true}
\usepackage[page]{appendix}
\usepackage{caption}
\usepackage{subcaption}
\usepackage{graphicx}
\usepackage{algorithm}
\usepackage{algpseudocode}
\usepackage{fancybox,framed}
\usepackage{multirow}
\usepackage{tabularx}
\usepackage{soul}
\usepackage{amsmath}
\usepackage{enumitem}
\usepackage{multirow}
\usepackage{diagbox}
\usepackage[export]{adjustbox}
\usepackage{array}

\title{Utilizing TTS Synthesized Data for Efficient Development of Keyword Spotting Model}

\name[affiliation={1}]{Hyun Jin}{Park}
\name[affiliation={1}]{Dhruuv }{Agarwal}
\name[affiliation={1}]{Neng}{Chen}
\name[affiliation={1}]{Rentao}{Sun}
\name[affiliation={1}]{Kurt}{Partridge}
\name[affiliation={1}]{Justin}{Chen}
\name[affiliation={1}]{Harry}{Zhang}
\name[affiliation={1}]{Pai}{Zhu}
\name[affiliation={1}]{Jacob}{Bartel}
\name[affiliation={1}]{Kyle}{Kastner}
\name[affiliation={1}]{Gary}{Wang}
\name[affiliation={1}]{Andrew}{Rosenberg}
\name[affiliation={1}]{Quan}{Wang}

\address{
  $^1$Google LLC, Mountain View, CA, U.S.A.}

\email{\{hjpark, dhruuv, nengchen,  sunrentao, kep, jstchen, harryz, paizhu, bartel, kkastner, wgary, rosenberg, quanw\}@google.com}

\keywords{keyword spotting, TTS synthesized training data}

\begin{document}

\maketitle

\begin{abstract}
    
  This paper explores the use of TTS synthesized training data for KWS (keyword spotting) task while minimizing development cost and time. Keyword spotting models require a huge amount of training data to be accurate, and obtaining such training data can be costly. In the current state of the art, TTS models can generate large amounts of natural-sounding data, which can help reducing cost and time for KWS model development. Still, TTS generated data can be lacking diversity compared to real data. To pursue maximizing KWS model accuracy under the constraint of limited resources and current TTS capability, we explored various strategies to mix TTS data and real human speech data, with a focus on minimizing real data use and maximizing diversity of TTS output. Our experimental results indicate that relatively small amounts of real audio data with speaker diversity (100 speakers, 2k utterances) and large amounts of TTS synthesized data can achieve reasonably high accuracy (within 3x error rate of baseline), compared to the baseline (trained with 3.8M real positive utterances).
\end{abstract}

\section{Introduction}
\label{sec:introduction}

The keyword spotting (KWS) task is to detect spoken keywords while rejecting background speech and noise. KWS has become an important mechanism for activating conversational human-computer interfaces since the advances in ASR. Representative examples include virtual assistants like Alexa, Siri, and Google Assistant, where KWS technology is used to start user-assistant interaction~\cite{AlexaMulti16, HeySiri17, Alvarez2019}.

A production level KWS system must cover a huge variety of conditions due to the diversity of populations, pronunciations, and acoustic environments. Also, keyword spotting models should be “always on”---ideally strictly causal for low latency, and with a small computational footprint to limit energy consumption.
To meet these requirements, there has been substantial KWS research using neural networks. Prior work has shown significant quality improvement and latency reduction in
low-resource settings~\cite{AlexaMulti16, HeySiri17, Alvarez2019,small_fp_kws, comp_td_nn_kws, MaxPool20, cascade_kws, He2018StreamingES, Cascade17, Custom17, Cnn15, Alexa16}.

Production level KWS models are usually trained with large amounts of training data to cover the wide diversity of pronunciation and acoustic environments. Gathering audio data specific to a target keyword incurs significant cost, as it requires human contributors to generate audio recordings.

Recent advancements in TTS (Text To Speech) allow generation of realistic audio at low cost. This has inspired many applications of generated data for the ASR domain~\cite{Xue2022ImprovingSR, Chen2020ImprovingSR, Deng2021ImprovingRF,Huang2023TextGW, Yang2023TextIA,rosenberg2019speech,tts4pretrain2}. For ASR, TTS models enable the use of text-only data, which is much more  abundant than labeled audio. This benefit often leads to improved accuracy and reduced data cost.

Following the success in ASR, there have been efforts to utilize TTS-generated data for KWS~\cite{Lin2020TrainingKS, Werchniak2021ExploringTA}. Lin et al proposed to pre-train an embedding model with real data, and fine-tune attached classifier head models using limited amounts of TTS or real data~\cite{Lin2020TrainingKS}. Werchniak et al showed initial exploration of TTS data usage for single keyword detection problem ~\cite{Werchniak2021ExploringTA} where mixing real and TTS data gave the best results.

The current state of the art TTS models can generate large amounts of synthetic data that sound natural to the human ear, but the generated data distribution may not match the distribution of real data. To address this mismatch, we propose TTS-based KWS model training strategies, based on three key components. 

Firstly we develop a text generator that generates text phrases tailored for KWS training. The text generator is designed to maximize diversity of TTS synthesized output. Secondly, we utilize state of the art TTS modules that can synthesize speech with large number of voices. The TTS modules provide a large number of pre-trained voices, and support the generation of personalized voices based on input audio. 

Finally, we evaluate various strategies to mix synthetic TTS data and real human speech data, with a focus of minimizing data cost while maximizing quality. To minimize cost and time for KWS model development, we evaluated mixing options that uses as small amount of real data as possible, while using large amounts of TTS generated data.

The contributions of this paper are : (1) We explore KWS model training using large amounts of synthetic data and \emph{minimum amounts of real data} to achieve comparable accuracy to the baseline which uses large scale real positive data. (2) We also provide \emph{reports on trade off relationship} between amount of used real data and model accuracy in multiple sweep conditions. (3) We propose a text generator that creates TTS input texts tailored to maximize diversity of TTS output by utilizing experimental \emph{prosody control} feature of Virtuoso TTS (section \ref{sec:text_augmentation}).

\section{Related works}
\label{sec:relatedworks}

For improved data diversity, we use two different TTS models. 

\subsection{Virtuoso TTS}
\label{sec:virtuoso_tts}

Virtuoso is a multilingual speech-text joint training model that can learn from untranscribed speech, unspoken text, and paired speech-text data sources~\cite{Saeki2022VirtuosoMM, Saeki2024ExtendingMS}. This model is capable of generating speech in 139 languages for 726 predefined speaker profiles. We used the simple text-to-speech mode of the Virtuoso model. Given a transcript, the model can generate an utterance for a target language from a designated speaker, with randomized prosody. Experiments with Virtuoso TTS model indicates that punctuation symbols in the text input can be used to control the prosody of the synthesized speech, and we leverage this capability of the Virtuoso model to augment our synthesized data.

\subsection{AudioLM TTS}

AudioLM is an audio generative language model that features long term coherence and high quality~\cite{Borsos2022AudioLMAL}. We used a variant of AudioLM-based TTS model that can be conditioned by text and audio, with the key feature of synthesizing audio while retaining the speaker's characteristics and prosody of the input audio~\cite{speartts}. The diversity of the generated dataset is able to match a rich variety of real human audio prompts.

\section{Baseline keyword spotting model}
\label{sec:kws_baseline}

\subsection{Input features}
In this study, we followed the lead of previous research~\cite{Alvarez2019, MaxPool20} and employed the same input features. Specifically, we extracted a 40-dimensional vector of spectral filterbank energies (calculated over a 25-millisecond window) at every 10-millisecond time frame. To create a 120-dimensional input feature vector $X_t$ for every 20 milliseconds, we stacked and strided three consecutive frames.
To enhance the model's resilience and adaptability, we incorporated data augmentation techniques. We followed the approach in~\cite{kim2017generation}, applying established methods like simulated reverberation and noise mixing to the data before feature extraction.

\subsection{Architecture}
\label{sec:architecture}

\begin{figure}
	\centering
	\includegraphics[width=\columnwidth]{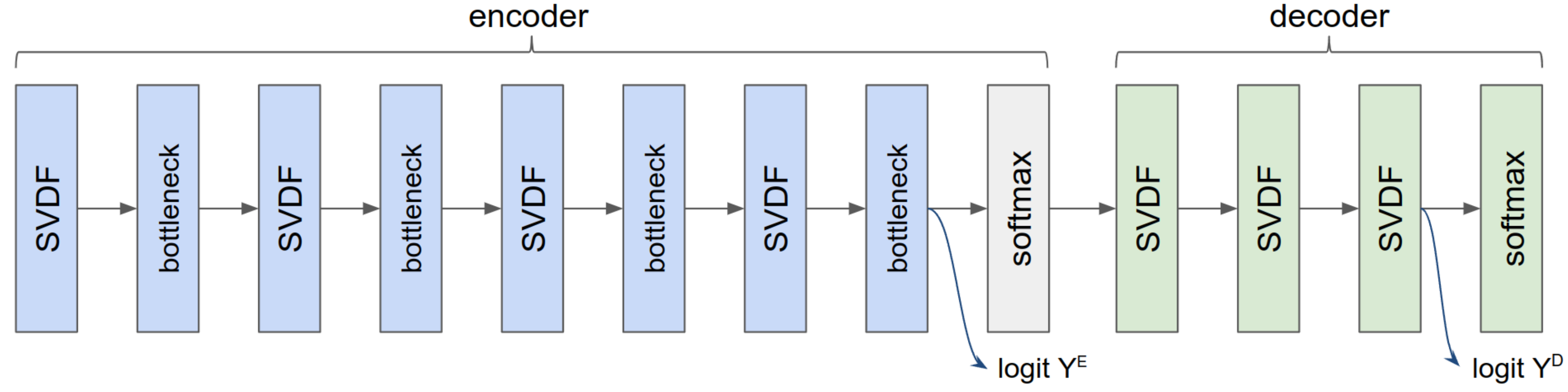}
	\caption{Baseline KWS model architecture}
	\label{fig:architecture}
\end{figure}

We employed a two-stage model architecture (Fig.~\ref{fig:architecture}), as outlined in previous work~\cite{Alvarez2019, MaxPool20}. This architecture comprises seven factored convolution layers (also known as SVDF~\cite{Alvarez2019}) and three bottleneck projection layers, organized into encoder and decoder sub-modules connected sequentially. The model is optimized for streaming inference, and has roughly 320,000 parameters.

The encoder module takes as input $X_t$, a vector of stacked spectral filter-bank energies representing the audio features. It then produces an $N$-dimensional encoder output $Y^\textrm{E}$, which is designed to encapsulate $N$ phoneme-like sound units crucial for keyword recognition. This encoded representation is then passed to the decoder module, which generates a 2-dimensional output $Y^\textrm{D}$ trained to predict the presence or absence of the keyword within the audio stream. The final prediction logit, denoted as Y, combines both the encoder and decoder outputs: $Y=[Y^\textrm{E},Y^\textrm{D}]$. This unified representation enables robust keyword spotting in diverse audio environments.

\subsection{Supervised training objective}
\label{sec:supervised}

The baseline training approach leverages two types of supervised losses (Eq.~\ref{eq:mlmp-loss}). The first loss term directly calculates the cross-entropy between model logits and labels, following the method established in~\cite{Alvarez2019}. The second loss term computes the cross-entropy between max-pooled logits and labels, as introduced in~\cite{MaxPool20}.  Both loss terms have distinct components for the encoder and decoder modules, and a weighted combination of these components forms the final loss.

The top level loss is a weighted combination of the cross-entropy and max-pooled losses (Eq.~\ref{eq:mlmp-loss}). This combination helps prevent overfitting and improves the model's ability to perform well on unseen data, ultimately enhancing its robustness and effectiveness in keyword spotting tasks.

\begin{equation}
\label{eq:mlmp-loss}
\begin{split}
\mathcal{L}_{\textrm{sup}} = \sum_{t=1..n} [&(1 - \alpha) L_{\textrm{CE}}\left (Y(X_t,\theta), c_t  \right ) \\
 &+ \alpha L_{\textrm{MP}}\left (Y(X_t,\theta), \omega_{\textrm{end}}  \right )]
\end{split}
\end{equation}

$Y(X_t,\theta)$ denotes the combined encoder and decoder model output given input $X_t$ and parameter set $\theta$. 
$L_{\textrm{CE}}$ represents the end-to-end loss proposed in Alvarez~\cite{Alvarez2019}. 
We use the implementation as defined by Eq. 2 in~Park et al.\cite{MaxPool20}  where $c_t$ is the per-frame target label for CE-loss. 
$L_{\textrm{MP}}$ represents the max-pool loss proposed in~Park et al.\cite{MaxPool20}, which was defined by Eq. 12.
$\omega_{\textrm{end}}$ represents the end-of-keyword position label for the max-pool loss. $\alpha$ is a loss-weighting hyper-parameter determined empirically. Refer to~\cite{Alvarez2019, MaxPool20} for details of $L_{\textrm{MP}}$ and $L_{\textrm{CE}}$.

\section{Proposed approach}
\label{sec:proposed_method}

\begin{figure}
	\centering
	\includegraphics[width=\columnwidth]{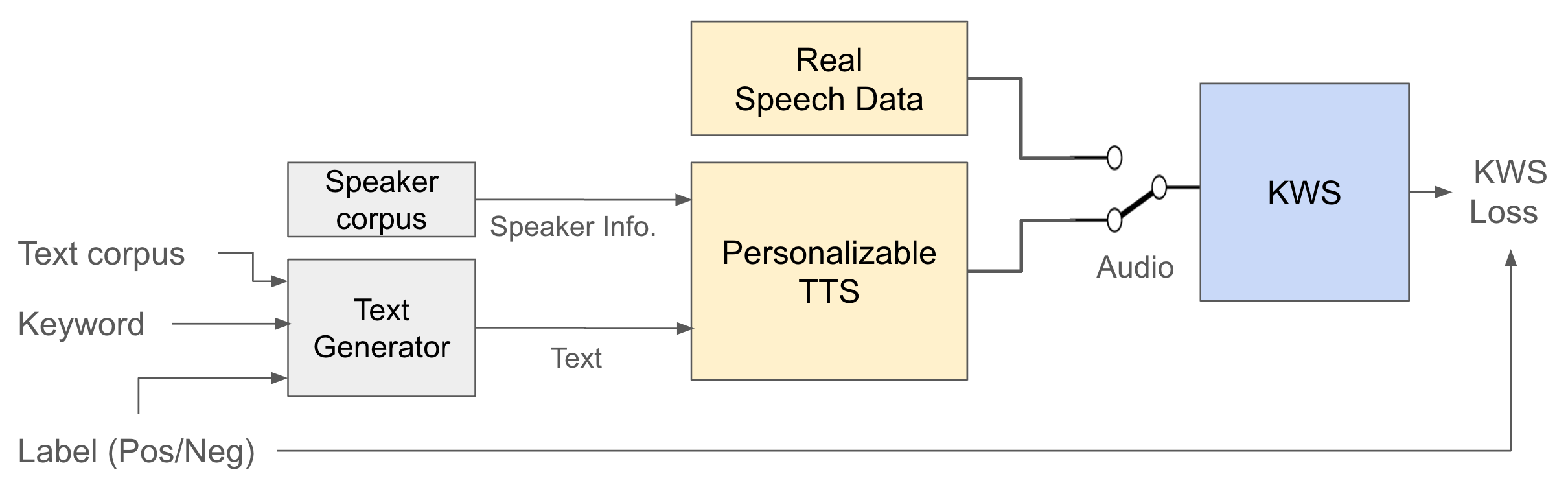}
	\caption{ Proposed KWS approach with TTS generated data }
	\label{fig:tts_hw_approach}
\end{figure}

Fig.~\ref{fig:tts_hw_approach} shows the high level view of the proposed TTS based KWS training approach. In this approach, the baseline KWS model can take input audio examples from either real speech data or TTS generated data sources. We explore training recipes that mix both real and synthesized data. The mixing ratio is a hyper-parameter which we explore in sweep experiments. We use TTS models that can generate speech samples in various speaker types and locales. The TTS models are conditioned by both speaker information and text. Speaker information can be either an index of the prefixed speaker (Virtuoso), or audio samples from any speaker (AudioLM). Text input to TTS is generated by a text generator, which combines target keyword and randomized negative text from a text corpus depending on target label (positive or negative). Details of the major components of the proposed approach are discussed below.

\subsection{Text augmentation by text generator}
\label{sec:text_augmentation}

We introduces a text generator module, which generates positive or negative phrase examples given target label (pos/neg), target keywords (for example, ``hey google'' or ``ok google'') 
and a random text corpus (which constitutes negative phrases in the form of user queries following the keyword).

We define a keyword as a combination of $prefix$ and $key\_name$ (keyword := ($prefix$, $key\_name$)). For example $prefix$ can be "Hey" and $key\_name$ can be "Google". Based on given keyword (prefix and key\_name) and random query text, we build a positive phrase by concatenating. Negative phrases can be constituted by using any textual corpus and filter out the keyword. We also randomly add some prosody control symbols (Table~\ref{tab:prosody_controls}) to vary the TTS output. Those prosody control symbols are experimental features of Virtuoso TTS model. We first define a set of text templates with variables (prefix, key\_name, and query) and prosody control symbols (Table~\ref{tab:text_templates}). Variables can be replaced with actual provided texts. Then those text templates  are randomly sampled to generate TTS input texts.

\begin{table}[h!]
	\begin{center}
		\caption{Prosody control symbols}
		\label{tab:prosody_controls}
		\begin{tabular}{c|c}
		\hline
		Controlled texts & Effects  \\
		\hline
		$text$   & default pronunciation of $text$ \\
		($text$) & speak $text$ slowly \\
		$text$: & insert pause after $text$ \\
		$text$? & increase pitch at the end of $text$ \\
		$text$! & speak $text$ loudly \\
		\hline
		\end{tabular}
	\end{center}
\end{table}

\begin{table}[h!]
	\begin{center}
		\caption{Text generation templates}
		\label{tab:text_templates}
		\begin{tabular}{c|c}
		\hline
		Text templates & Notes  \\
		\hline
    \{$prefix$\} \{$key\_name$\} \{$query$\}     & positive phrase \\
    \{$prefix$\} (\{$key\_name$\}) \{$query$\}   & positive phrase \\
    (\{$prefix$\}): (\{$key\_name$\}) \{$query$\} & positive phrase \\
    \{$prefix$\}: (\{$key\_name$\})? \{$query$\}  & positive phrase \\
    \{$prefix$\}: \{$key\_name$\}! \{$query$\}    & positive phrase \\
    \{$query$\} & negative phrase \\
		\hline
		\end{tabular}
	\end{center}
\end{table}

\subsection{TTS data generation by Personalizable TTS}

The TTS models used in our approach provide capabilities to generate diverse and personalized speech. The Virtuoso system supports 726 pretrained high quality speakers. AudioLM-based TTS model can generate speech with a voice matching with the input audio example. We maximally utilize such personalization capability of the TTS to generate training examples with diverse voices.

Also Virtuoso is highly multilingual model supporting 139 languages. We also utilize this feature to generate speech with different language targets from fixed English phrases. The resulting TTS output sounds like an accented version of English, adding diversity to the synthesized output.

\subsection{TTS and Real data mixing strategy}

Although current state of the art TTS can generate realistic human speech, the distribution of TTS generated data and real speech data might mismatch. TTS generated data might still have artifacts that does not exist in real speech recordings, and it might not generate all the variations present in real speech. To compensate for such mismatch we explore the strategy of mixing real data with TTS based data, by training models with various data mixing options, and evaluate them on real data.

\section{Experimental Setup}
\label{sec:experiment_setup}

\subsection{Input Data}
We trained KWS models for "Hey/OK Google" detection task using real and synthesized data. For real speech data, we used anonymized utterances  collected in accordance to Google's Privacy and AI Principles \cite{privacyprinciples, aiprinciples}. TTS data is generated using Virtuoso and AudioLM variant TTS models. Multi-style data augmentation \cite{KimMTR2017} is applied during training. Table \ref{tab:training_data} summarizes the number of utterances used. TTS data numbers include both Virtuoso and AudioLM sources. 

\begin{table}[h!]
	\begin{center}
		\caption{Data types and sizes}
		\label{tab:training_data}
		\begin{tabular}{l|r}
		\hline
		Data Types &  Utterance counts  \\
		\hline
Real Positive Utts &		3.8 M \hspace{0.6cm} \\
Real Negative Utts &		14.1 M \hspace{0.6cm}  \\
		\hline
Synthesized Positive Utts 	&	7.5 M \hspace{0.6cm}  \\
Synthesized Negative Utts	&	5.1 M \hspace{0.6cm}  \\
		\hline
		\end{tabular}
        \vspace*{-4mm}	
	\end{center}
\end{table}

\subsection{Evaluation Data and Metric}
\label{sec:data_and_metric}

We evaluate the model performance on real Hey/OK Google data sets. Our primary metric to compare model performance is false reject rate (FRR) and the model threshold is selected to optimize FRR while keeping the maximum allowed false accept per hour fixed at $0.133$ which is a typical operational condition.

\subsection{Different experiment sweeps}

\begin{itemize}

\item Baseline model performance on different data-sets

In this experiment, we aim to establish the baseline metric scores when the model is trained on different data-sets including synthesized and real data.

We also explore addition of real negative data. Obtaining positive data, is constrained by the selected the key-word(s) while negative data can be obtained from virtually any data source as long as it doesn't contain the target keyword. Hence, we explored adding such negative sets, here-by called as \emph{base} real negative data, to improve baseline performance.

\item TTS with incremental amounts of real positive data

In this experiment we wish to understand the need of real data for model training when quality TTS data is available. To do so, we compare performance of pure TTS trained model, and gradually increase the real positive data to 100k. We also test the effect of addition of real \emph{base} negatives here, mentioned above, and see how it can offset the need for real data drastically.

\item TTS data and varying amount of speakers in real data

We train models with the baseline of fixed TTS configuration, while adding real data uniformly per speaker and gradually increasing speaker count. Uniformly sampling real utterances from speakers should provide data-diversity to help model train better and faster ideally. 

\item TTS data and varying number of utterances per speaker in real data

Similar to the above experiment, we keep the TTS data as fixed, but now vary the number of utterances per speaker while keeping the speaker count as fixed. Instead of increasing diversity of speakers, we aim to see how many utterances per speaker helps the model.



\end{itemize}

\section{Experimental Results}
\label{sec:experiment_setup}

\subsection{Baseline KWS models with simple mixing options}

Table \ref{tab:baseline_small_model} shows the evaluation results of models trained with different combinations of training data. We show FRR's at a fixed false accept rate (section \ref{sec:data_and_metric}). TTS data trained FRR's are high (46.47\%) compared to real data trained baseline FRR (3.17\%).

However, we observe that adding real negative data ($\sim$11 M) to all sweep experiments improves FRR numbers of all TTS based models dramatically (the second half of Table \ref{tab:baseline_small_model}). Based on the observation and, we decide to keep using real negative data as \emph{base} negative data for all other experiments. We also see that mixing all TTS and real data gives the best results.

\begin{table}[h!]
	\begin{center}
        \vspace*{-1mm}	
		\caption{Model baselines trained on different datasets}
		\label{tab:baseline_small_model}
		\begin{tabular}{l|r}
		\hline
		\textbf{Baseline models per train data} &  {FRR} \\
		\hline
Virtuoso data only &		53.10\% \\
AudioLM data only &		46.50\% \\
TTS (Virtuoso + AudioLM)  &		46.47\% \\
Real data only  &		3.17\% \\
		\hline
        \textbf{Improved baselines models per train data} &  {FRR}  \\
		\hline
Virtuoso + \textit{base real negative data} &		17.75\% \hspace{0.0cm} \\
AudioLM + \textit{base real negative data} &		16.59\% \hspace{0.0cm} \\
TTS  + \textit{base real negative data} &		17.94\% \hspace{0.0cm} \\
TTS + Real data &		2.46\% \hspace{0.0cm} \\
		\hline
		\end{tabular}
        \vspace*{-3mm}	
	\end{center}
\end{table}

\subsection{TTS + Incremental amounts of real positive data}

\begin{figure}
	\centering
	\includegraphics[width=\columnwidth]{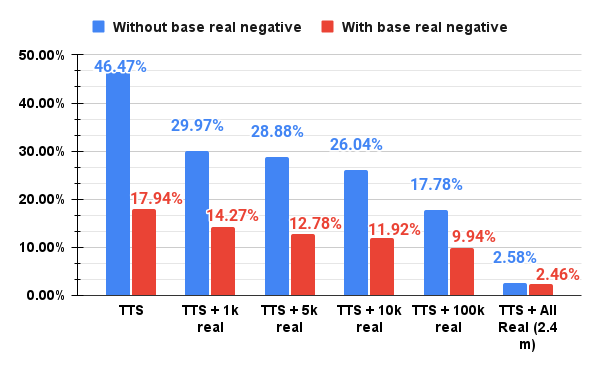}
	\caption{FRR over different amounts of real data. Blue bars indicate the baseline configs, and red bars have additional \emph{base} negative data. Medium sized model used here.) }
	\label{fig:datasweep_logs_label}
\end{figure}

In this setup, we train models with all the TTS data and gradually increasing amounts of real positive data (randomly sampled) as shown in Fig. \ref{fig:datasweep_logs_label}. The blue bars show the FRR numbers of the models trained without real negative data, while the red bars show FRR numbers of models trained with \emph{base} real negative data. Blue and Red bars show similar improvements when real negative data is added as \emph{base} training data.

In Fig. \ref{fig:datasweep_logs_label} we see that the model FRR's improves (decreases) monotonically, as we add more real \emph{positive} data to the training data. From the Fig., we can observe that all the real negative data improved FRR number from 46.7\% down to 17.94\% with TTS only baseline. On top of that adding all the real positive data to TTS only  baseline improved FRR from 17.94\% to 2.46\%. We can conclude that both negative and positive real data have a significant impact to the TTS only baseline model. And we note that about 100k of real positive data samples with \emph{base} real negative data and TTS data gives 9.94\% FRR, which is about $\sim$3 times the FRR of real data only baseline (3.17\%).

\subsection{TTS + Real data with varying number of speakers}
\label{sec:TTS_Real_speaker_count_variation}

In this setup, we use all the TTS data with the real negative data as the base training data, and mixes small amounts of real positive data that has increasing number of speakers. We sample fixed number (10) of utterances per each speaker. The first half of Table \ref{tab:speakers_sweep_medium_model} shows the result, where the FRR numbers improves as the number of speakers in the real positive data grows.
We observe that models trained with number of speakers 100 or higher, gives  performance with FRR similar to or less than 3 times the FRR of the baseline (3.17\%). Note that the baseline model is trained with 3.8M real positive data, while the TTS+100 speaker model was trained with only 1k real positive utterances.

\begin{table}[h!]
	\begin{center}
		\caption{Speaker variation experiments. First we increase speaker count while fixed utterances per speaker to 10, and second, we fix the speaker count to 100, and increase utterances per speaker. All exp uses full TTS data and base negative data.}
		\label{tab:speakers_sweep_medium_model}
		\begin{tabular}{l|r}

		\hline
		\textbf{Speakers with 10 utterances each} &  {FRR} \hspace{0.2cm} \\
		\hline
TTS + 1 Speaker \hspace{6mm}(10 utts)  &		15.28\% \hspace{0.0cm} \\
TTS + 10 Speakers \hspace{2.5mm} (100 utts)  &		14.94\% \hspace{0.0cm} \\
TTS + 100 Speakers \hspace{2mm}(1k utts)   &		9.78\% \hspace{0.0cm} \\
TTS + 200 Speakers \hspace{2mm}(2k utts)  &		9.90\% \hspace{0.0cm} \\
TTS + 500 Speakers \hspace{2mm}(5k utts)  &		\textbf{7.63}\% \hspace{0.0cm} \\
		\hline
		\textbf{Speaker count fixed at 100} &  {FRR} \hspace{0.2cm} \\
		\hline
TTS + 2 utts/speaker \hspace{3mm}(200 utts)  &		10.99\% \hspace{0.0cm} \\
TTS + 6 utts/speaker \hspace{3mm}(600 utts) &		10.95\% \hspace{0.0cm} \\
TTS + 12 utts/speaker \hspace{1.5mm}(1.2k utts)   &		10.71\% \hspace{0.0cm} \\
TTS + 20 utts/speaker \hspace{1mm} (2k utts)  &		9.47\% \hspace{0.0cm} \\
TTS + 200 utts/speaker (20k utts)  &		7.99\% \hspace{0.0cm} \\
		\hline
		\end{tabular}
	\end{center}
\end{table}

\subsection{TTS + Real data with varying number of utterances per speaker}

As a complimentary experiment to section \ref{sec:TTS_Real_speaker_count_variation}, we keep the speaker count constant (100) and increase the number of utterances per speaker with the real positive data. Second half of Table \ref{tab:speakers_sweep_medium_model}, shows that increasing number of utterances improves FRR relatively \emph{slowly} given fixed number of speakers. From this observation, we can conclude that increasing diversity of speakers as the previous sweep has more impact than increasing number of utterances with fixed speaker count.

\section{Conclusion}
We explore and evaluate the benefits of TTS synthesized data for training keyword spotting models. To maximize diversity of generated data, we applied prosody controlled text generation, and used TTS generation conditioned by audio samples. Finally we evaluated various mixing conditions of real data that ranges from zero to a scale, where we evaluated the effects of number of speakers, and number of utterances per speakers. Experiments showed that we are able to obtain reasonably high accuracy ($\sim$ 3 times FRR of baseline) with a fractional number of real positive samples (2k) compared to the baseline (3.8M). 

\section{Acknowledgements}
\label{sec:acknowledgements}

The authors would like to acknowledge the support from Charles Yoon, Pedro Meningbar, Bhuvana Ramabhadran, and Fran{\c{c}}oise~Beaufays.

\pagebreak

\bibliographystyle{IEEEtran}
\bibliography{template}

\end{document}